\documentclass[a4paper,11pt]{article}

\usepackage{amsmath,amssymb,bbm}
\usepackage{cancel}
\usepackage[dvipdfm]{graphicx}
\usepackage[dvipdfm]{hyperref}
\usepackage{ulem}
\hypersetup{
pdftitle={Worldsheet Instanton Corrections to 522-brane Geometry},
pdfauthor={Tetsuji Kimura and Shin Sasaki},
colorlinks={true},
linkcolor={black},
urlcolor={black},
filecolor={black},
citecolor={blue}
}
\makeatletter

 \@addtoreset{equation}{section}
\makeatother

\newcounter{Enumerate}

\DeclareFontFamily{U}{rsf}{}
\DeclareFontShape{U}{rsf}{m}{n}{
  <5> <6> rsfs5 <7> <8> <9> rsfs7 <10-> rsfs10}{}
\DeclareMathAlphabet\Scr{U}{rsf}{m}{n}
\usepackage[mathscr]{eucal}

\newcommand{\Vev}[1]{\big{\langle} {#1} \big{\rangle}}

\newcommand{\del}{\partial}

\newcommand{\half}{\frac{1}{2}}

\newcommand{\LS}{\ \ \ \ \ \ \ \ \ \ }
\newcommand{\ls}{\ \ \ \ \ }
\newcommand{\wt}{\widetilde}
\newcommand{\wh}{\widehat}
\newcommand{\ve}{\varepsilon}
\newcommand{\ol}{\overline}

\newcommand{\bsubeq}{\begin{subequations}}
\newcommand{\esubeq}{\end{subequations}}
\newcommand{\noi}{\noindent}

\newcommand{\nn}{\nonumber}

\newcommand{\N}{\mathcal{N}}

\renewcommand{\d}{{\rm d}}
\newcommand{\e}{{\rm e}}
\renewcommand{\i}{{\rm i}}

\newcommand{\slb}{\scalebox}

\def\+{{+\!\!\!+}} 

\begin{document}
\allowdisplaybreaks{

\thispagestyle{empty}

%%%%%%%%% title %%%%%%%%%%%%%

\begin{flushright}
RUP-13-5
\\
\end{flushright}

\vspace{30mm}

\noi
\slb{2.5}{Worldsheet Instanton Corrections to}

\vspace{3mm}

\noi
\slb{2.5}{$5^2_2$-brane Geometry}

\vspace{7mm}

\noi
{\renewcommand{\arraystretch}{1.7}
\begin{tabular}{cl}
\multicolumn{2}{l}{\slb{1.0}{Tetsuji {\sc Kimura}}}
\\
& \slb{.9}{\renewcommand{\arraystretch}{1.0}
\begin{tabular}{l}
{\sl
Department of Physics and Research Center for Mathematical Physics,
Rikkyo University} 
\\
{\sl Tokyo 171-8501, JAPAN
}
\\
\slb{0.9}{\tt tetsuji \_at\_ rikkyo.ac.jp}
\end{tabular}
}
\\
& \multicolumn{1}{l}{and}
\\
\multicolumn{2}{l}{\slb{1.0}{Shin {\sc Sasaki}}}
\\
& {\renewcommand{\arraystretch}{1.0}
\begin{tabular}{l}
{\sl
Department of Physics,
Kitasato University}
\\
{\sl Sagamihara 252-0373, JAPAN}
\\
\slb{0.9}{\tt shin-s \_at\_ kitasato-u.ac.jp}
\end{tabular}
}
\end{tabular}
}

\vspace{14mm}

%%%%%%%%% abstract %%%%%%%%
\noindent
\slb{1.1}{\sc Abstract}:
\begin{center}
\slb{.95}{
\begin{minipage}{.9\textwidth}
We study worldsheet instanton corrections to the 
exotic $5^2_2$-brane geometry in type II string theory.
The BPS vortices in the $\mathcal{N} = (4,4)$ 
gauged linear sigma model 
modify the geometry of the $5^2_2$-brane.
We find that the modification of the geometry is 
understood by the localization in the
 T-dualized winding direction.
\end{minipage}
}
\end{center}

%%%%%%%%%%%%%%%%%%%%%%%%%%%%%%%%%%%%%%%%%%%%%%%%%%%%%%%%%%%%%%%%%%%%%%
\newpage
\section{Introduction}

In string theory, the Buscher rule of T-duality \cite{Buscher:1987sk} 
exhibits the mixing of the metric and the NSNS B-field 
of spacetime geometries.
Performing T-duality transformations 
along several directions,
one obtains a non-conventional object
whose metric is no longer single-valued.
Due to the pathological nature, these kind of objects in string theory
have been ignored for a long time.
Recently, the non-conventional geometries again appear 
in string compactification scenarios with non-vanishing background fields. 
One of the most familiar 
example is known as the T-fold \cite{Hull:2004in}.
This contains information of the general coordinate transformation and the T-duality transformation.
The feature of the T-fold 
comes from the winding modes as well as the momentum modes 
of the string coupled to the B-field 
on the space.
Since the contribution of the winding modes cannot be represented in the conventional geometry, 
they are refereed to as the ``non-geometric'' backgrounds.
A typical example of the T-fold is the exotic $5_2^2$-brane 
\cite{deBoer:2012ma}.
One finds this less familiar object from NS5-brane in type II string
theory via T-duality along two transverse directions.
So far, this has been well investigated in the supergravity viewpoint 
\cite{deBoer:2012ma, Kikuchi:2012za, Hassler:2013wsa}.

A salient feature of stringy corrections to the geometry comes from
the worldsheet quantum effects. 
The gauged linear sigma model (GLSM) is a powerful tool to study phases of vacua 
\cite{Mirror03}
and quantum corrections to string worldsheet theory 
\cite{Morrison:1994fr}. 
The IR limit of the GLSM is 
the nonlinear sigma model (NLSM) whose target space
is a solution to supergravities.
The GLSM is known to be useful 
because of the following reasons:
(i) the gauge multiplets in the UV regime connects various phases of vacua in the system, and
(ii) the soliton solutions provided by the gauge fields can be mapped to
the worldsheet instantons in the IR regime.
In the previous paper \cite{Kimura:2013fda}, we have constructed the
$\N=(4,4)$ GLSM for five-branes of co-dimension three
which gives rise to
the nonlinear sigma model whose target space is the exotic
$5^2_2$-brane. 
The model is obtained from the GLSMs for the NS5-branes and for the
Kaluza-Klein (KK) monopoles via T-dualities. 
In \cite{Tong:2002rq, Harvey:2005ab, Okuyama:2005gx}, 
the instanton corrections to the NS5-branes and KK-monopoles are studied
through the vortices in the GLSMs.
In the present paper, 
based on the GLSM constructed in \cite{Kimura:2013fda}, 
we study the instanton corrections to the geometry
of the exotic $5^2_2$-brane.

The organization of this paper is as follows:
In Section \ref{Tdual-chains}
we briefly review the T-duality chain among the NS5-branes, 
the KK-monopoles and the exotic $5_2^2$-brane in the viewpoint of supergravity.
We also discuss the instanton corrections and their geometrical meaning to the former two branes.
In Section \ref{inst-522} 
we study the worldsheet instanton corrections to the exotic $5_2^2$-brane from the two distinct perspectives.
One is from the T-dualized description of the KK-monopoles involving the instanton corrections.
The other is from the 
direct calculation of instantons in the GLSM.
We find that the instanton corrections to the geometry of the co-dimension three
five-branes are carried over to that of the co-dimension two exotic
$5^2_2$-brane. 
Although the two results are obtained from the different routes, we find
that they are completely consistent with each other.
Section \ref{sect-conclusion} is devoted to conclusion and discussions.

%%%%%%%%%%%%%%%%%%%%%%%%%%%%%%%%%%%%%%%%%%%%%%%%%%%%%%%%%%%%%%%%%%%%%%
\section{T-duality chain among NS5-branes, KK-monopoles and exotic
 $5^2_2$-brane} \label{Tdual-chains}
In this section, we briefly explain the T-duality chain among 
the NS5-branes, the KK-monopoles and the exotic $5^2_2$-brane in
type II string theory. 
These objects are realized as solutions to the equations of motion 
in ten-dimensional supergravities. 
Throughout this paper, we consider the configurations of the objects 
depicted in Table \ref{tb:NS5-KKM-522}.
%%%%%%%%%%%%%%%%%%%%%%%%%%%%%%%
\begin{table}[htbp]
\begin{center}
\begin{tabular}{|l|cccccccccc|}
\hline
 & 0 & 1 & 2 & 3 & 4 & 5 & 6 & 7 & 8 & 9 \\
\hline \hline
NS5-branes & \textbullet & & & \textbullet & \textbullet & \textbullet &
							 \textbullet & \textbullet & 
 & \\
\hline
KK-monopoles & \textbullet & & & \textbullet & \textbullet & \textbullet &
							 \textbullet & \textbullet &  & $\circ$ \\
\hline
$5^2_2$-brane & \textbullet & & & \textbullet & \textbullet & \textbullet &
							 \textbullet & \textbullet & $\circ$ &
										 $\circ$ \\
\hline
\end{tabular}
\end{center}
\caption{The black dot \textbullet \ stands for the world-volume directions of each
 object. The others are transverse directions. In particular, starting from
 the NS5-branes, the compact directions where the T-duality 
transformations are performed are shown by the symbol $\circ$.}
\label{tb:NS5-KKM-522}
\end{table}
%%%%%%%%%%%%%%%%%%%%%%%%%%%%%%%
These geometries are realized as the target spaces of the string
NLSM. The action of the general supersymmetric NLSM is given by
\cite{Gates:1984nk},
\begin{align}
S \ = \ 
\frac{1}{2\pi} \int \! \d^2 x \ 
&\left[
- \frac{1}{2} g_{\mu \nu} \, \partial_{m} X^{\mu} \, \partial^m X^{\nu} 
+ \frac{1}{2} B_{\mu \nu} \, \varepsilon^{mn} \, \partial_m X^{\mu} \, 
\partial_n X^{\nu}
\right. 
\nn \\
\ & \ \ 
\left.
+ \frac{\i}{2} g_{\mu \nu} \, \Omega^{\mu}_{-} D_{+} \Omega^{\nu}_{-} 
+ \frac{\i}{2} g_{\mu \nu} \, \Omega^{\mu}_{+} D_{-} \Omega^{\nu}_{+}
+ \frac{1}{4} R_{\mu \nu \rho \sigma} \, \Omega^{\mu}_{+} \Omega^{\nu}_{+}
 \Omega^{\rho}_{-} \Omega^{\sigma}_{-}
\right]
\, ,
\label{eq:general_NLSM}
\end{align}
where $D_{\pm} = D^{\pm}_0 \pm D^{\pm}_1$, $\varepsilon^{01} = 1$ and
the worldsheet metric is $\eta_{mn} = \mathrm{diag} (-1,1)$. 
The covariant derivative $D^{\pm}_m$ is defined with the positive 
and negative torsion. 
The functions $g_{\mu \nu} (X)$ and $B_{\mu \nu} (X)$ are 
the target space metric and the anti-symmetric NSNS B-field.
The function $R_{\mu \nu \rho \sigma}$ is the Riemann tensor of the
target space with the positive torsion. 
The real fermions $\Omega^{\mu}_{\pm}$ are supersymmetric
partners of the string coordinates $X^{\mu}$.

It is known that the string quantum corrections 
modify the target space geometries \cite{Mirror03}.
In particular, we focus on the worldsheet instanton corrections to the
 geometries. 
The worldsheet instantons 
are best examined in the language of the GLSM 
\cite{Mirror03, Morrison:1994fr} whose IR limit describes the NLSM
\eqref{eq:general_NLSM}. 
As we will discuss in the following section, 
instanton corrections to the four-point function of
fermions in the GLSM in the UV regime result in the corrections to
the Riemann tensor in the low-energy effective action
\eqref{eq:general_NLSM}. 
The instanton effects on the geometries are extracted from the Riemann
tensor $R_{\mu \nu \rho \sigma}$. 
In the following subsections we introduce the worldsheet
instanton corrections to the geometries of the H- and KK-monopoles.

\subsection{NS5-branes and H-monopoles}
Our T-duality chain starts from the NS5-branes in type II string theory.
The NS5-branes are known to be the co-dimension four solitonic solution 
to ten-dimensional supergravity equations of motion \cite{Callan:1991dj}.  
The explicit form of the solution is given by
\bsubeq \label{eq:CHS}
\begin{align}
\d s^2_{\text{NS5}} \ &= \ 
\d x^2_{034567} + H (\vec{R}) \, \d x^2_{1289}
\, , \quad 
\e^{2 \phi} \ = \ H(\vec{R})
\, , \\
H_{\mu \nu \rho} \ &= \ 
\epsilon_{\mu \nu \rho} {}^{\lambda}
 \partial_{\lambda} \log H (\vec{R})
\, , \quad 
H (\vec{R}) \ = \ 
\frac{1}{g^2} + \frac{Q}{|\vec{R}|^2}
\, ,
\end{align}
\esubeq
where $\phi$ is the dilaton and $H_{\mu \nu \rho} \ (\mu,\nu,\rho = 1,2,
8,9)$ is the field strength
of the NSNS B-field. The constant $g$ is a dimensionless parameter and 
$Q$ is a charge with appropriate mass dimension. The vector $\vec{R} =
(X^1, X^2, X^8, X^9)$ represents the coordinates in the $\mathbb{R}^4$ plane. 
We used the notation such as
$\d x^2_{034567} = - (\d X^0)^2 + (\d X^3)^2 + \cdots + (\d X^7)^2$
and 
$\d x^2_{1289} = (\d X^1)^2 + (\d X^2)^2 + (\d X^8)^2 + (\d X^9)^2 $.
Now we compactify the $X^9$-direction on $S^1$ with 
radius $\mathcal{R}_9$. Then the harmonic function $H$ becomes
\begin{align}
H (r, \vartheta) \ &= \ 
\frac{1}{g^2} + \sum_{n=-\infty}^{\infty} 
\frac{Q}{r^2 + (\vartheta - 2 \pi \mathcal{R}_9 n)^2}
\, , \label{eq:X9sum}
\end{align}
where $\vec{r} = (X^1, X^2, X^8)$, $r = |\vec{r}|$ and $\vartheta = X^9$.
The summation in \eqref{eq:X9sum} is performed in the small 
$\mathcal{R}_9$ limit.
In this limit, the discrete summation over $n$ 
is approximated by the continuous integral. The result is independent of $\vartheta$:
\begin{align}
H (r,\vartheta) \ &= \ 
\frac{1}{g^2} 
+  \int_{-\infty}^{\infty} \! \d n \
\frac{Q}{r^2 + (\vartheta - 2 \pi \mathcal{R}_9 n)^2}
= \frac{1}{g^2} + \frac{Q \mathcal{R}_9^{-1}}{2 r}
\, .
\label{eq:H-monopole}
\end{align}
The solution \eqref{eq:CHS} with the harmonic function 
\eqref{eq:H-monopole} is called 
the H-monopoles or the smeared NS5-branes,
i.e., the H-monopoles are obtained by compactifying a transverse direction in
the NS5-brane solution \eqref{eq:CHS}. 
The H-monopole solution possesses 
a $U(1)$ isometry along the $X^9$-direction.
We note that the asymptotic radius of the $S^1$ direction is characterized by $1/g$.

On the other hand, when 
the radius ${\cal R}_9$ is kept finite
in the summation \eqref{eq:X9sum}, the result becomes
\begin{align}
H (r,\vartheta) \ = \ 
\frac{1}{g^2} + \frac{Q \mathcal{R}_9^{-1}}{2 r} \frac{\sinh (r
 \mathcal{R}_9^{-1})}{\cosh (r \mathcal{R}^{-1}_9) - \cos (\vartheta \mathcal{R}^{-1}_9)}
\, .
\label{eq:localized_H}
\end{align}
The solution \eqref{eq:CHS} with the harmonic function
\eqref{eq:localized_H} is called the localized H-monopoles \cite{Gregory:1997te}. 
In this case, the isometry along the $X^9$-direction is broken. 
Although the relation between the solutions with \eqref{eq:H-monopole} and
\eqref{eq:localized_H} is obvious in the viewpoint of supergravity, it
is interesting to observe these results in the worldsheet perspective.
The expansion of the function
\begin{align}
\frac{\sinh r}{\cosh r - \cos \vartheta} 
\ = \ 
1 + \sum_{n=1}^{\infty} (\e^{- n r + \i n \vartheta} + \e^{- n r - \i n \vartheta})
\end{align}
strongly suggests contributions to the harmonic function
\eqref{eq:H-monopole} from topological sectors labeled by the
integer $n$ behind the effect of the finite radius $\mathcal{R}_9$.
Indeed, it is pointed out in \cite{Tong:2002rq} 
that the integer $n$ is nothing but the 
topological winding number stemming from the string worldsheet
instantons.
The worldsheet instantons are
interpreted as the BPS vortices in the $\mathcal{N} = (4,4)$ 
GLSM whose IR limit describes the H-monopoles.
It is demonstrated that all the $n$ instantons 
contribute to the four-point function of fermions and modify the geometry of the
H-monopoles.
The calculations are performed in the large asymptotic radius limit $1/g \to \infty$. 
It is explicitly shown that the instanton corrections break the isometry along
$X^9$ and precisely reproduce the harmonic
function \eqref{eq:localized_H} of the localized H-monopoles.
From now on we refer to 
this kind of instantons
as the $X^9$-instantons.

\subsection{KK-monopoles}
The KK-monopoles are known to be 
T-dual of the H-monopoles.
For the H-monopoles, the solution has an isometry along
the $X^9$-direction. Therefore 
we can perform the T-duality transformation along 
this direction. Applying the 
Buscher rule \cite{Buscher:1987sk} to the geometry \eqref{eq:CHS} with
\eqref{eq:H-monopole}, 
we have the KK-monopole geometry,
\bsubeq \label{eq:KKM}
\begin{gather}
\d s^2_{\text{KKM}} \ = \ 
\d x^2_{034567} + H (r) \, \d x^2_{128} 
+ H^{-1} (r) \big( \d 
\wt{X}^9 + \tfrac{1}{2} \omega_i \, \d X^i \big)^2 
\quad 
(i=1,2,8),  \\
\d \omega \ = \ *_3 \d H
\, , \quad 
\phi \ = \ 0
\, , \quad 
H (r) \ = \ 
\frac{1}{g^2} + \frac{\wt{\mathcal{R}}_9}{2r}
\, , \quad 
\wt{\mathcal{R}}_9 \ = \ Q \mathcal{R}_9^{-1}
\, ,
\end{gather}
\esubeq
where $\wt{X}^9$ is the dual coordinate of $X^9$ in the H-monopoles. 
The Hodge dual $*_3$ is defined in $\mathbb{R}^3$ spanned by $(X^1,X^2,X^8)$.
Then the vector $\vec{\omega}$ satisfies the Dirac monopole
equation, $\mathrm{rot} \, \vec{\omega} = - 2 \vec{\nabla} H$. 
The transverse directions to the KK-monopoles are the Taub-NUT space
which possesses a $U(1)$ isometry. 
In the asymptotic regime $r \to \infty$, the transverse space is the
non-trivial fibration of $S^1$ over $\mathbb{R}^3$. 
The asymptotic radius of the $S^1$ direction is characterized by $g$, 
which is just the T-dual circle. 
It is natural to consider the worldsheet instanton
 corrections\footnote{We call this also the $X^9$-instantons not $\wt{X}^9$-instantons.}
to the KK-monopole geometry \eqref{eq:KKM}. 
In \cite{Harvey:2005ab}, the instanton effects in the GLSM for the
KK-monopoles are studied. 
As in the same way in the H-monopoles, the BPS vortices in the GLSM modify
the target space geometry \eqref{eq:KKM}. 
The instanton corrections are interpreted as the modification of the
harmonic function $H(r)$ in \eqref{eq:KKM}. The authors in \cite{Harvey:2005ab}
found that the modified function is nothing but the one in \eqref{eq:localized_H} 
obtained in the localized H-monopoles. 
Then the harmonic function depends on the ``T-dual coordinate'' $\vartheta$, i.e., 
the KK-monopole is localized along the {\it winding} direction. 
Therefore the solution \eqref{eq:KKM} with the harmonic function
\eqref{eq:localized_H} is called the localized KK-monopoles.

A few comments are in order. 
First, the harmonic function depending on the winding coordinate $\vartheta$ 
rather than the physical coordinate $\wt{X}^9$ 
indicates the non-geometric structure of the solution.
This kind of non-geometric nature is studied in the doubled formalism
 \cite{Hull:2006va, Jensen:2011jna}.
Second, the configuration \eqref{eq:KKM} with the modified harmonic
function \eqref{eq:localized_H} is no longer the solution to the BPS
equations in supergravity.
This is because the calculation is
performed in the small radius limit $g \to 0$ of the T-dual circle
\cite{Harvey:2005ab}. In the small radius limit, the KK modes become massive
while the string winding modes become lighter. 
We therefore need to incorporate the light winding modes into supergravities.
Then the supergravity approximation of 
the string massless spectrum is lost. 
Finally, we note that even though the localized KK-monopoles show the non-geometric nature, 
they also preserve a $U(1)$ isometry along the 
$X^8$-direction when we further compactify this direction on $S^1$ 
(see Table \ref{tb:NS5-KKM-522}). This fact 
enables us to T-dualize the KK-monopoles with 
the $X^9$-instanton corrections 
into the $5_2^2$-brane geometry.
We will discuss this issue in Section \ref{inst-522}.

\subsection{Exotic $5^2_2$-brane}
We next proceed to the exotic $5^2_2$-brane.
The geometry of the $5^2_2$-brane is obtained by performing the T-duality 
transformation on the KK-monopoles.
In order to utilize the 
Buscher rule, we compactify the 
$X^8$-direction on $S^1$ with radius $\mathcal{R}_8$ in the KK-monopole geometry
\eqref{eq:KKM}. 
The harmonic function becomes
\begin{align}
H \ = \ 
\frac{1}{g^2} + \frac{1}{2} \sum_{l = - \infty}^{\infty} 
\frac{\wt{\mathcal{R}}_9}{\sqrt{\varrho^2 + (X^8 - 2 \pi \mathcal{R}_8 l)^2}}
\, , \quad 
\varrho^2 = (X^1)^2 + (X^2)^2
\, .
\end{align}
We approximate the discrete summation over $l$ by the 
continuous integration in the small $\mathcal{R}_8$ \cite{deBoer:2012ma}. 
Since the integration over $l$ diverges, we introduce the cutoff
$\Lambda$ to regularize it. 
The result is 
\begin{align}
H \ = \ 
h_0 + \frac{\wt{\mathcal{R}}_9}{2 \pi \mathcal{R}_8} \log \frac{\mu}{\varrho}
\, , \label{eq:522h}
\end{align}
where $\mu > 0$ is the renormalization scale and 
$h_0 = \frac{1}{g^2} +
\frac{\wt{\mathcal{R}}_9}{2 \pi \mathcal{R}_8} \log \frac{4 \pi \Lambda}{\mu}$
is a constant which
diverges in the limit $\Lambda \to \infty$. 
It is obvious that the harmonic function is ill-defined for 
$\varrho > \mu$. 
Therefore the metric is well-defined only in the region 
$\varrho < \mu$. 
The physical meaning of this renormalization scale is discussed in
 \cite{Kikuchi:2012za}.
The vector $\vec{\omega} = (\omega_1, \omega_2, \omega_8)$ is obtained by solving the Dirac
monopole equation with the harmonic function \eqref{eq:522h}.
Using the Buscher rule on the geometry \eqref{eq:KKM}, we find the $5^2_2$-brane solution,
\bsubeq \label{eq:522}
\begin{gather}
\d s^2_{5^2_2} \ = \ 
\d x^2_{034567} 
+ H \, \d x^2_{12} + H K^{-1} \big( (\d\wt{X}^8)^2 + (\d \wt{X}^9)^2 \big)
\, , \\
B_{\wt{8}\wt{9}} \ = \ \frac{1}{2} K^{-1} \omega_8
\, , \quad 
K \ = \ H^2 + \frac{1}{4} \omega_8^2
\, , \quad 
\e^{2 \phi} \ = \ H K^{-1}
\, , \quad 
\omega_8 \ = \ 
\frac{\wt{\mathcal{R}}_9 \wt{\mathcal{R}}_8}{\pi \alpha'} \arctan \left(
\frac{X^2}{X^1}
\right)
\, ,
\end{gather}
\esubeq
where we have employed the gauge where $\omega_{1} = \omega_2 = 0$ and
$\wt{\mathcal{R}}_8 = \alpha'/\mathcal{R}_8$ is the dual radius of
$\mathcal{R}_8$. 
The harmonic function is given by \eqref{eq:522h}. 
We point out that the asymptotic radius of the second T-dual circle
along $X^8$-direction, 
which is not the $X^9$-isometry direction in the Taub-NUT, 
cannot be seen in this geometry 
because of the IR renormalization scale $\varrho < \mu$. 
This is true even in the GLSM which will be discussed in Section 3.
We can not introduce an extra parameter which represents the asymptotic
radius of the second T-dual circle.

The geometry \eqref{eq:522} shows some exotic nature. 
For example, due to the function $\mathrm{arctan} (X^2/X^1)$, the
metric is not single valued. Therefore it has non-trivial 
monodromy around the origin in the $(X^1, X^2)$-plane. 
Indeed, this is a common property of exotic objects \cite{deBoer:2012ma}. 
Even though the $5^2_2$-brane is
exotic, it is straightforward to realize the geometry \eqref{eq:522} 
in the string nonlinear sigma model. 
In the next section, we examine 
the $X^9$-instanton corrections to the $5^2_2$-brane geometry.

%%%%%%%%%%%%%%%%%%%%%%%%%%%%%%%%%%%%%%%%%%%%%%%%%%%%%%%%%%%%%%%%%%%%%%
\section{Instanton corrections to $5^2_2$-brane geometry}
\label{inst-522}
In this section, we study the worldsheet instanton corrections to the
exotic $5^2_2$-brane geometry in two independent ways. 
The first is through the localized KK-monopoles.
We calculate the corrections to the $5^2_2$-brane geometry from the
viewpoint of the $X^9$-instantons in the KK-monopole geometry.
We then perform the T-duality transformation along the $X^8$-direction by using the 
Buscher rule.
Notice that the $X^9$-instanton corrections to the KK-monopoles 
\cite{Harvey:2005ab} preserve the isometry along 
the $X^8$-direction, whilst the one along the $X^9$-direction is broken.
The second is through the multi-centered five-branes.
In the previous paper, we have studied the $\mathcal{N} = (4,4)$ GLSM 
 which gives rise to the NLSM with the target space geometry for
 co-dimension three multi-centered $k$ five-branes.
In the following discussions we refer to them as $\wh{5}^2_2$-branes.
The NLSM for the co-dimension two $5^2_2$-brane is obtained 
from the $\wh{5}^2_2$-branes by the suitable choice of
the Fayet-Iliopoulos (FI) parameters and the limit $k \to \infty$ 
\cite{Kimura:2013fda}. 
We evaluate the instanton corrections to the 
four-point function of fermions in the GLSM. 
The instanton corrections modify the 
geometry of the $\wh{5}^2_2$-branes in the IR limit. 
The modified geometry is reduced to that of the co-dimension two 
$5^2_2$-brane by the above procedure.
We will show that the two results coincide with each
other even though those are obtained from the different ways.

\subsection{KK-monopole perspective}
We have discussed the 
$X^9$-instanton corrections to the geometry of the
KK-monopoles in Section \ref{Tdual-chains}. 
We observed that the harmonic function of the KK-monopole is modified
and is given in \eqref{eq:localized_H}.
Now we compactify the $X^8$-direction on $S^1$ with radius $\mathcal{R}_8$.
The localized KK-monopoles become those of the periodic array with 
an infinite number of mirror images. The harmonic function \eqref{eq:localized_H} becomes
\begin{align}
H \ = \ 
\frac{1}{g^2} + \frac{1}{2} 
\sum_{l = -\infty}^{\infty} 
\frac{\wt{\mathcal{R}}_9}{\sqrt{\varrho^2 + (X^8 - 2 \pi \mathcal{R}_8 l)^2}}
\left[
\frac{\sinh (\mathcal{R}_9^{-1} \sqrt{\varrho^2 + 
(X^8 - 2 \pi \mathcal{R}_8 l)^2})}{\cosh
(\mathcal{R}_9^{-1} \sqrt{\varrho^2 + (X^8 - 2 \pi \mathcal{R}_8 l)^2}) 
- \cos (\mathcal{R}_9^{-1} \vartheta)}
\right]
\, .
\label{eq:X8X9sum}
\end{align}
Since the summation over $l$ diverges, we need to regularize it.
In order to find the regularization, 
we first decompose the harmonic function as the summation over the 
$n$ $X^9$-instantons: 
\begin{align}
H \ &= \ 
\frac{1}{g^2} 
+ \sum_{l = -\infty}^{\infty} 
\frac{\wt{\mathcal{R}}_9}{\sqrt{\varrho^2 + (X^8 - 2 \pi \mathcal{R}_8 l)^2}}
\nn \\
\ & \ \ \ \ 
\times \left[
1
+
\sum_{n=1}^{\infty}
\left(
\e^{- n \mathcal{R}^{-1}_9 \sqrt{\varrho^2 + (X^8 - 2\pi \mathcal{R}_8
 l)^2} + \i n \mathcal{R}^{-1}_9 \vartheta}
+
\e^{- n \mathcal{R}^{-1}_9 \sqrt{\varrho^2 + (X^8 - 2\pi \mathcal{R}_8
 l)^2} - \i n \mathcal{R}^{-1}_9 \vartheta}
\right)
\right]
\, .
\end{align}
We approximate the discrete summation over $l$ by
the continuous integration in the small $\mathcal{R}_8$,
\begin{align}
H \ = \ 
\frac{1}{g^2} + \frac{\wt{\mathcal{R}}_9}{2\pi {\cal R}_8} \int^{\infty}_{0} \! \d l \
\frac{1}{\varrho^2 + l^2}
+ \frac{\wt{\mathcal{R}}_9}{2\pi {\cal R}_8} 
\sum_{n = 1}^{\infty} 
\left(
\e^{\i n \mathcal{R}_9^{-1} \vartheta} + \e^{- \i n \mathcal{R}_9^{-1} \vartheta}
\right)
\int^{\infty}_{0} \! \d l \ \frac{1}{\varrho^2 + l^2} 
\e^{- n \sqrt{\varrho^2 + l^2}}
\, .
\end{align}
The second term diverges while the third term remains finite. 
The second term together with the first term 
has been already calculated in \eqref{eq:522h}. 
The third term becomes 
\begin{align}
\frac{\wt{\mathcal{R}}_9}{2\pi {\cal R}_8} 
\sum_{n = 1}^{\infty} 
\left(
\e^{\i n \mathcal{R}_9^{-1} \vartheta} + \e^{- \i n \mathcal{R}_9^{-1} \vartheta}
\right) K_0 (n \mathcal{R}^{-1}_9 \varrho)
\, ,
\end{align}
where $K_{0} (x)$ is the modified Bessel function of the second kind. 
This harmonic function does not break the isometry along 
the $X^8$-direction.
Therefore we can perform the T-duality transformation along 
this direction. Applying the Buscher rule, 
we obtain the $5^2_2$-brane geometry with the harmonic function given by 
\begin{align}
H \ = \ 
h_0 + \frac{\wt{\mathcal{R}}_9 \wt{\mathcal{R}}_8}{2\pi \alpha'} \log \frac{\mu}{\varrho} 
+ \frac{\wt{\mathcal{R}}_9 \wt{\mathcal{R}}_8}{2\pi \alpha'} 
\sum_{n \neq 0}
\e^{\i n \mathcal{R}_9^{-1} \vartheta} K_0 (|n| \mathcal{R}^{-1}_9 \varrho)
\, . \label{eq:522harmonic}
\end{align}
The first and second terms are just the harmonic function of the 
$5^2_2$-brane \eqref{eq:522h}. The third term is corrections coming from the 
$X^9$-instantons in the KK-monopoles. 
It is worth pointing out that the function \eqref{eq:522harmonic} 
also appears in the study of the D-instanton corrections to the 
moduli space of hypermultiplets in Calabi-Yau compactifications
\cite{Ooguri:1996me}. It is discussed that the singularity in the moduli space
metric is smoothed out by the D-instanton effects. 
Indeed, the function \eqref{eq:522harmonic} is regular
at $\varrho = 0$ provided that $\vartheta$ is non-zero. 
The distinct origin of the functional form \eqref{eq:522harmonic}
provides the geometric intuition behind this fact.
The $5^2_2$-brane is localized in the winding direction 
$\vartheta$, namely at the origin $\vartheta = 0$. 
The regularity of the metric at $\varrho = 0$
is interpreted due to the non-zero interval from the $5^2_2$-brane in the winding
space. The singularity appears again when one approaches to the
$5^2_2$-brane not only in the physical geometry 
$\varrho \to 0$ but also in the winding space $\vartheta \to 0$.

%%%%%%%%%%%%%%%%%%%%%%%%%%%%%%%%%%%%%%%%%%%%%%%%%%%%%%%%%%%%%%%%%%%%%%
\subsection{$\wh{5}^2_2$-branes perspective}
In this subsection, we study 
the instanton corrections to the $5^2_2$-brane geometry 
from the direct calculations by the GLSM.
In \cite{Kimura:2013fda}, we have studied the
$\mathcal{N} = (4,4)$ GLSM for 
the co-dimension three multi-centered $k$ $\wh{5}^2_2$-branes.
Starting from the GLSM for the H-monopoles 
and performing the T-duality transformation, we obtain the GLSM for 
 the KK-monopoles.
Taking once again the T-dual along the other direction, we obtain the 
$\mathcal{N} = (4,4)$ GLSM for the $\wh{5}^2_2$-branes.
The bosonic sector of the GLSM Lagrangian for the $\wh{5}^2_2$-branes 
is 
\begin{align}
\Scr{L}
\ &= \ 
\sum_{a=1}^k \frac{1}{e^2_a} 
\Big\{
\frac{1}{2} (F_{01,a})^2 - | \partial_m \sigma_a |^2 
- 4 | \partial_m M_{c,a} |^2 
\Big\}
\nn \\
\ & \ \ \ \ 
- \frac{1}{2 g^2} 
\Big\{
(\partial_m r^1)^2 + (\partial_m r^3)^2 
\Big\}
- \frac{g^2}{2} 
\Big\{
(\partial_m y^2)^2 + (D_m \gamma^4)^2
\Big\} 
\nn \\
\ & \ \ \ \ 
- \sum_{a=1}^k 
\Big\{
| D_m q_a |^2 + | D_m \wt{q}_a |^2 
\Big\}
- \sqrt{2} \, \varepsilon^{mn} \sum_{a = 1}^k 
\partial_m 
\left(
(\vartheta - t_{2,a}) A_{n,a}
\right) 
\nn \\
\ & \ \ \ \ 
- 2 g^2 \sum_{a,b = 1}^k 
\left(
\sigma_a \ol{\sigma}_b + 4 M_{c,a} \ol{M}_{c,b}
\right)
- 2 \sum_{a=1}^k 
\left(
|\sigma_a|^2 + 4 |M_{c,a}|^2
\right) 
(|q_a|^2 + |\wt{q}_a|^2)
\nn \\
\ & \ \ \ \ 
- \sum_{a=1}^k 
\frac{e^2_a}{2} 
\Big(
|q_a|^2 - |\wt{q}_a|^2 - \sqrt{2} \, (r^3 - t_{1,a})^2 
\Big)^2
- \sum_{a=1}^k e^2_{a} 
\Big| \sqrt{2} \, q_a \wt{q}_a 
- \left(
(r^1 - s_{1,a}) + \i (r^2 - s_{2,a})
\right)
\Big|^2 
\nn \\
\ & \ \ \ \ 
+ \frac{g^2}{2} \sum_{a,b = 1}^k
\left(
A_{c=,a} + \ol{A}_{c=,a}
\right)
\left(
B_{c\+, b} + \ol{B}_{c\+,b}
\right)
\, .
\label{eq:522GLSM}
\end{align}
The model exhibits $U(1)^k$ gauge 
symmetries with gauge fields $A_{m,a}$. 
The complex scalar fields $q_a$ and $\wt{q}_a$ are components in the $\N=(4,4)$ charged hypermultiplets.
The real scalar fields $r^1$, $r^2$, $r^3$ and $\gamma^4$ belong to 
the $\N=(4,4)$ neutral hypermultiplet.
%Under the duality transformations \cite{Rocek:1991ps}, 
%the scalar fields $y^2$ and $\vartheta$ are related to $r^2$ and
%$\gamma^4$, respectively. 
The $\N=(4,4)$ gauge multiplets contain various scalar fields:
$\sigma_a$ and $M_{c,a}$ are complex scalar fields. 
$A_{c=,a}$, $\ol{A}_{c=,a}$, $B_{c\+, b}$ and $\ol{B}_{c\+,b}$ are auxiliary fields.
%(for the details, see \cite{Kimura:2013fda}).
The gauge coupling constants $e_a$ are 
dimensionful while $g$ is dimensionless. 
The real FI parameters $t_{1,a}, s_{1,a}, s_{2,a}$ represent positions of the $k$ 
$\wh{5}^2_2$-branes  in the $(r^1,r^2,r^3)$-directions 
while $t_{2,a}$ are shifts in the $\vartheta$-direction.

The model contains the kinetic terms for 
the scalar fields $r^1, r^3, y^2, \gamma^4, \sigma_a, M_{c,a}, q_a,
\wt{q}_a$ and the gauge fields $A_{m,a}$. 
From the duality transformation \cite{Rocek:1991ps} between the
H-monopoles and the KK-monopoles, 
the field $\gamma^4$ is an implicit function of $\vartheta$ and
$A_{m,a}$ through the following relation
\cite{Harvey:2005ab},
\begin{align}
\pm (\partial_0 \pm \partial_1) \vartheta = 
- g^2 
\left\{
(\partial_0 \pm \partial_1) \gamma^4 + \sqrt{2} 
\sum_{a=1}^k (A_{0,a} \pm A_{1,a})
\right\}, 
\end{align}
where the field $\vartheta$ is identified
with the $X^9$ coordinate in the H-monopoles.
Under the duality transformations between the 
KK-monopoles and the $\wh{5}^2_2$-branes, 
the scalar field $r^2$ is given as an implicit function of $y^2$ and the
auxiliary fields $A_{c=,a}$, $B_{c\+, b}$ through the relations \cite{Kimura:2013fda},
\bsubeq
\begin{align}
 (\partial_0 + \partial_1) r^2 =& 
- g^2 (\partial_0 + \partial_1) y^2 
+ g^2 \sum_{a=1}^k (B_{c\+, a} + \ol{B}_{c\+,a}), \\
 (\partial_0 - \partial_1) r^2 =& 
 g^2 (\partial_0 - \partial_1) y^2 
+ g^2 \sum_{a=1}^k (A_{c=, a} + \ol{A}_{c=,a}).
\end{align}
\esubeq
The fields $\sigma_a, M_{c,a}, A_{m,a}$ become auxiliary fields in the
IR limit and do not have geometrical meaning.
The charged fields $q_a$, $\wt{q}_a$ give constrains on the fields $r^1,
r^2, r^3$ in the IR.
Therfeore the independent geometric coordinates of the
$\wh{5}^2_2$-branes are given by $r^1, r^3, y^2$ and $\gamma^4$ which
will be redefined as $\wt{\vartheta}$ (see the equation
\eqref{eq:gamma4} in below).
The classical supersymmetric vacua of the GLSM 
in the $a$-th sector are found from the potential terms.
From the positive definite parts, we find $\sigma_a = M_{c,a} = 0$. 
We take the branch where the 
last term in \eqref{eq:522GLSM} vanishes.
The fields $(r^1, r^2, r^3)$ and $(q_a, \wt{q}_a)$ are the 
triplet and the doublet in the $SU(2)$ rotation. 
We employ a frame where 
$q_a \neq 0$,
$\wt{q}_a = 0$ and $r^1 = s_{1,a}$, $r^2 = s_{2,a}$ in the $a$-th sector.
The vacuum expectation value of $r^3$ becomes a modulus of the $a$-th
vacuum $r^3 = \zeta_a$.
From these conditions, we have $|q_a| = \sqrt{2} (\zeta_a - t_{1,a})$.
We have also $y^2 = \gamma^4 = 0$, and $A_{m,a}$ stay at the pure gauge.

The IR limit of the GLSM \eqref{eq:522GLSM} is obtained by
taking $e_a \to \infty$. 
In this limit, the fields $A_{m,a}$,
$\sigma_a$, $M_{c,a}$ are frozen and become auxiliary fields.
From the potential terms, 
we obtain constraints among fields \cite{Kimura:2013fda}.
By using the constraints and integrating out all the auxiliary fields,
we obtain the nonlinear sigma model given by
\begin{align}
\Scr{L}
\ &= \ 
- \frac{1}{2} H \Big\{ (\del_m r^1)^2 + (\del_m r^2)^2 + (\del_m r^3)^2 \Big\}
- \sqrt{2} \, \ve^{mn} 
\sum_{a=1}^k
\del_m ((\vartheta - t_{2,a}) {A}_{n,a})
+ \ve^{mn} (\del_m r^2) (\del_n y^2) 
\nn \\
\ & \ \ \ \ 
- \frac{1}{2} H^{-1} (\del_m \wt{\vartheta})^2 
- \half ({\omega}_{2})^2 H^{-1} (\del_m r^2)^2
%-
+ {\omega}_{2} H^{-1} 
(\del_m \wt{\vartheta}) (\del^m r^2) 
\nn \\
\ & \ \ \ \ 
- \half ({\omega}_{1})^2 H^{-1} (\del_m r^1)^2
- {\omega}_{1} {\omega}_{2} H^{-1} (\del_m r^1) (\del^m r^2)
%- 
+{\omega}_{1} H^{-1} (\del_m \wt{\vartheta}) (\del^m r^1)
\, , \label{eq:522NLSM}
\end{align}
where we have rewritten the field $\gamma^4$ by $\wt{\vartheta}$ as 
\begin{align}
\wt{\vartheta} = \gamma^4 
%- 
+ \sqrt{2} \sum_{a=1}^k \alpha_a.
\label{eq:gamma4}
\end{align}
Here $\alpha_a$ are the phase factors of $q_a$ and $\wt{q}_a$.
The function $H$ is defined by 
\begin{align}
H \ = \ 
\frac{1}{g^2} + \sum_{a = 1}^k \frac{1}{\sqrt{2} R_a}
\, , \quad 
R_a \ = \ 
\sqrt{(r^1 - s_{1,a})^2 + (r^2 - s_{2,a})^2 + (r^3 - t_{1,a})^2}
\, ,
\label{eq:five-brane_h}
\end{align}
and the set $(\omega_1, \omega_2)$ 
is the solution to the Dirac monopole equation
with the harmonic function \eqref{eq:five-brane_h}.
The target space geometry of \eqref{eq:522NLSM} represents 
the multi-centered $k$ $\wh{5}^2_2$-branes of co-dimension three.
In order to obtain the $5_2^2$-brane of co-dimension two, 
we compactify the $r^2$-direction on $S^1$ 
with radius $\mathscr{R}$
\footnote{
This procedure is justified at the last stage of the 
T-dulaity transformation where $r^2$ is integrated out explicitly and
the metric of the $5^2_2$-brane is obtained correctly \cite{Kimura:2013fda}.}
. 
The FI parameters $s_{1,a}, t_{1,a}, t_{2,a}$ should  
be independent of $a$ in the compactification. 
In the following, we take $s_{1,a} = t_{1,a} = t_{2,a} = 0$ for simplicity.
The positions of the $\wh{5}^2_2$-branes 
 in the $r^2$-direction become 
 $s_{2,a} = 2 \pi \mathscr{R} a$, $a \in \mathbb{Z}$ and $k \to
 \infty$.
The summation over $a$ becomes continuous
integral in the small $\mathscr{R}$.
After integrating out the field $r^2$, 
the result is the nonlinear sigma model whose target space is the
$5^2_2$-brane geometry \eqref{eq:522} \cite{Kimura:2013fda}.

Now we are going to study 
the instanton corrections to the vacuum moduli
space in the GLSM \eqref{eq:522GLSM}.
As discussed in \cite{Tong:2002rq, Harvey:2005ab},
 there is no quantum moduli space
of vacua by the Coleman-Mermin-Wagner theorem \cite{Coleman:1973ci, Mermin:1966fe}.
However, we can study the quantum corrected moduli space of vacua in 
the IR regime by integrating out the high momentum modes.
It is also discussed that there is no exact BPS instanton solution in
the GLSM for the H- and KK-monopoles. 
The situation is the same even for the present case. 
Following \cite{Tong:2002rq, Harvey:2005ab}, we therefore consider constrained (point-like)
instantons \cite{Affleck:1980mp, Mirror03} 
in the limit $g \to 0$ where an exact BPS solution exists.
In the limit $g \to 0$, the fields $r^1$, $r^3$ are frozen and the kinetic
terms of $y^2$, $\gamma^4$ vanish.
In order to find the instantons in the $g \to 0$ limit, 
we consider the configuration where $A_{m,a}$, $q_a$ are 
dynamical fields and the others stay at the vacua.
Then we find the following truncated model:
\begin{align}
\Scr{L}
\ &= \ 
\sum_{a=1}^k \frac{1}{2 e^2_a} (F_{01,a})^2 
- \sum_{a=1}^k |D_m q_a|^2 
- \sum_{a=1}^k \frac{e^2_a}{2} 
\left(
|q_a|^2 - \sqrt{2} 
\zeta_a
\right)^2 
- \sqrt{2} \sum_{a = 1}^k 
\vartheta F_{01,a}
\, ,
\label{eq:truncated}
\end{align}
where we have dealt with the field $\vartheta$ as a constant.
This is a conceivable assumption since the field $\vartheta$ is no
longer dynamical and appears only in the topological term
in \eqref{eq:522GLSM}. 
This is a reminiscent of the instanton calculus in the KK-monopoles 
\cite{Harvey:2005ab, Okuyama:2005gx} where $\vartheta$ is recognized as
a constant parameter rather than the field on the worldsheet theory.
The truncated model \eqref{eq:truncated} 
is just the $k$ Abelian-Higgs model with 
the FI parameters $\zeta_a$.
After the Wick rotation to the Euclidean space ($x^2 = \i x^0$), we find
that the Lagrangian becomes
\begin{align}
\Scr{L}
_{\text{E}} 
\ &= \ 
\sum_{a=1}^k 
\left[
\frac{1}{2 e^2_a} (F_{12,a})^2 
+ |D_m q_a|^2 
+ \frac{e^2_a}{2} 
\left(
|q_a|^2 - \sqrt{2} \zeta_a
\right)^2 
+ \i \sqrt{2} \, \vartheta F_{12,a}
\right] 
\nn \\
\ &= \ 
\sum_{a=1}^k 
\left[
\frac{1}{2 e^2_{a}}
\left(
F_{12,a} \pm e_a^2 (|q_a|^2 - \sqrt{2} 
\zeta_a 
)
\right)^2
+
\big| (D_1 \pm \i D_2) q_a \big|^2
\pm \sqrt{2} 
\zeta_a F_{12,a}
\right. 
\nn \\
\ &\LS
 \left.
 \frac{}{}
+ 
\i \sqrt{2} \, 
\vartheta F_{12,a}
\right]
\, .
\end{align}
From the Bogomol'nyi bound, we find the BPS equations: 
\begin{align}
F_{12,a} \ = \ \mp e_a^2 
(|q_a|^2 - \sqrt{2} 
\zeta_a
)
\, , \ls
(D_1 \pm \i D_2) q_a = 0
\ls \text{with} \ \ \ a = 1, \cdots, k
\, .
\label{eq:BPSANO}
\end{align}
These are the Abrikosov-Nielsen-Olesen (ANO) vortex equations \cite{Abrikosov:1956sx}.
Then the action $S = \frac{1}{2\pi} \int \! \d^2 x \ \Scr{L}_{\mathrm{E}}$ becomes 
\begin{align}
S \ = \ 
\sqrt{2} \sum_{a = 1}^k 
\zeta_a |n_a|
- \i \sqrt{2} \vartheta  
\sum_{a = 1}^k 
n_a
\, .
\end{align}
Here we defined the instanton number in the $a$-th sector,
\begin{align}
n_a \ = \ 
- \frac{1}{2 \pi} \int \! \d^2 x \ F_{12,a}
\, .
\end{align}
The analytic solution to the equations \eqref{eq:BPSANO} is not known
but these satisfy the half BPS condition, namely, the solution keeps
$\mathcal{N} = (2,2)$ worldsheet supersymmetries.

The instanton contributions to the geometry are given by the sum of
those in each $U(1)$ sector \cite{Okuyama:2005gx}. 
The calculations are the same in each sector $a$.
For the time being, we concentrate on the instantons in the $a$-th sector.
The path integral on the $n_a$-instantons background 
is reduced to the integral over the instanton moduli space $\mathcal{M}_{n_a}$.
Since the instantons keep $\mathcal{N} = (2,2)$ supersymmetries, the
bosonic and fermionic non-zero modes of the solution are cancelled out. 
We only need the zero modes. 
The integral measure on the moduli space is discussed in \cite{Tong:2002rq}. 
We consider the four-point correlation function of 
$\psi_{a \pm}, \wt{\psi}_{a \pm}$, 
which are supersymmetric partners of the charged scalars $q_a$, $\wt{q}_a$, 
on the $n_a$-instantons background:
\begin{align}
G^{(n_a)}_4 (x_1,x_2,x_3,x_4) 
\ = \ 
\Vev{
\ol{\psi}_{a+} (x_1) \psi_{a-} (x_2) \wt{\psi}_{a +} (x_3)
 \ol{\wt{\psi}}{}_{a-} (x_4) 
}_{\text{$n_a$-instantons}}
\, . \label{eq:4pt}
\end{align}
For our purpose, we only need the asymptotic solution to the BPS
equations \eqref{eq:BPSANO} \cite{deVega:1976mi}.
The fermionic position moduli in the solutions $\psi_{a\pm}$, $\wt{\psi}_{a\pm}$ 
saturate the associated Grassmann measure in the moduli space integration.
Since the solution preserves $\mathcal{N} = (2,2)$ supersymmetries in two dimensions, 
the moduli action has corresponding supersymmetries in $0+0$
dimensions. Then there are terms with four centered fermionic moduli in
the moduli action \cite{deVega:1976mi} which saturate the Grassmann
measure associated with the centered fermionic moduli.
Therefore we conclude that all the $n_a$-instantons 
give the non-zero contributions in \eqref{eq:4pt} even though 
the metric on the centered moduli space is not known for $n_a > 1$.

We note that the constraints among the scalar fields $q_a$,
$\wt{q}_a$ and the neutral ones $r^1, r^2, r^3$ 
induce those on the supersymmetric partners of 
the charged fields $\psi_{a\pm}, \wt{\psi}_{a\pm}$ and the neutral fermions 
$\chi_{\pm}, \wt{\chi}_{\pm}$
(for the fermionic field contents, see \cite{Kimura:2013fda}),
\begin{align}
\chi_{\pm} \ = \ 
2 \sum_{a=1}^k ( \wt{q}_a \, \psi_{a\pm} + q_a \, \wt{\psi}_{a\pm})
\, , \quad 
\wt{\chi}_{\pm} 
\ = \ 2 \i \sum_{a=1}^k 
(\wt{q}_a \, \ol{\wt{\psi}}{}_{a\pm} - q_a \, \ol{\psi}{}_{a\pm})
\, .
\label{eq:fermion_relations}
\end{align}
The complex fermions $\chi_{\pm}$, $\wt{\chi}_{\pm}$ are decomposed
into the real fermions $\Omega^{\mu}_{\pm}$ \cite{Harvey:2005ab}. 
Therefore the four-point function \eqref{eq:4pt} is translated 
to the coefficient of the four-point interaction of $\Omega^{\mu}_{\pm}$ in the
IR Lagrangian \eqref{eq:general_NLSM}. 
Then we find the instanton corrections to the Riemann tensor of the 
target space geometry
for the $\wh{5}^2_2$-branes. 
Since the calculation itself is the same discussed in \cite{Tong:2002rq,
Harvey:2005ab}, we never repeat it here.
The corrections to the Riemann tensor 
at the leading order in $1/R_a$ in the limit $g\to 0$ is found to be
\bsubeq
\begin{align}
\delta R_{1313}
 \big|_a 
\ &= \
\delta R_{2323}
 \big|_a
\ = \ 
- \frac{\mathcal{N}}{4 R_a} 
\sum_{n_a=1}^{\infty} n_a^2 \, 
\e^{- n_a R_a \mathscr{R}^{\prime -1}
} 
\big( \e^{\i n_a \mathscr{R}^{\prime -1}
\vartheta} 
+ \e^{- \i n_a \mathscr{R}^{\prime -1} \vartheta} \big)
\, , \label{eq:Rg} \\
%%%%%%%%%%%%%%%%
\delta R_{1323}
 \big|_a 
\ &= \ 
- 
\delta R_{2313}
 \big|_a
\ = \ 
- \frac{\mathcal{N}}{4 R_a} 
\sum_{n_a=1}^{\infty} n_a^2 \, 
\e^{- n_a R_a \mathscr{R}^{\prime -1}} 
\big( \e^{\i n_a \mathscr{R}^{\prime -1} \vartheta} 
- \e^{- \i n_a \mathscr{R}^{\prime -1} \vartheta} \big)
\, , \label{eq:Rt}
\end{align}
\esubeq
where the symbol $|_a$ stands for the contributions from the $a$-th
instanton sector. 
The corrections \eqref{eq:Rg}, \eqref{eq:Rt} 
contain an overall factor $\mathcal{N}$
stemming from the volume of the centered
moduli space of the ANO vortices which no one knows
\footnote{
The factor $\sqrt{2}$ in the harmonic function in
\eqref{eq:five-brane_h} is also absorbed in this 
overall factor.
This factor $\sqrt{2}$ comes from the canonical normalization of the
two-dimensional scalar fields \cite{Kimura:2013fda} which is different from the one in \cite{Tong:2002rq}. 
}.
The dimensionful coordinates in \eqref{eq:Rg}, \eqref{eq:Rt} 
are normalized by a parameter $\mathscr{R}^{\prime}$.
The (anti)symmetric structure of the Riemann tensor suggests that 
\eqref{eq:Rg} comes from the corrections to the metric 
while \eqref{eq:Rt} from these to the torsion.
We note that the off-diagonal part of the target space metric in
\eqref{eq:522NLSM} are dropped in the limit $g \to 0$ and the non-zero
component of the metric is $g_{ij} = H \delta_{ij}$ 
where $i,j=1,2,3$.
Assuming that the instanton corrections are only in the diagonal part,
the result \eqref{eq:Rg} is reproduced by the following modification to 
the metric at the leading order in $1/R_a$ expansion
\cite{Tong:2002rq},
\begin{align}
\delta g_{11} \big|_a 
\ = \ \delta g_{22} \big|_a 
\ = \ \delta g_{33} \big|_a 
\ = \
\frac{\mathcal{N}}{2 R_a} 
\sum_{n_a=1}^{\infty} 
\e^{- n_a R_a \mathscr{R}^{\prime -1}} 
\big( \e^{\i n_a \mathscr{R}^{\prime -1}
\vartheta} 
+ \e^{- \i n_a \mathscr{R}^{\prime -1}
\vartheta} \big)
\, 
\, .
\end{align}
In addition, we have the corrections to the torsion
from \eqref{eq:Rt} \cite{Harvey:2005ab}, 
\begin{align}
T_{128} \big|_a 
\ &= \ 
- H_{128} \big|_a 
\ = \ 
- 
\frac{
i \mathcal{N} }{2 R_a} 
\sum_{n_a=1}^{\infty} 
n_a 
\e^{- n_a R_a \mathscr{R}^{\prime -1}} 
\big( \e^{\i n_a 
\mathscr{R}^{\prime -1} \vartheta} 
- \e^{- \i n_a \mathscr{R}^{\prime -1}
\vartheta} \big)
\, .
\end{align}
Therefore the instanton corrections to 
the $k$ $\wh{5}^2_2$-branes of co-dimension three 
\eqref{eq:522NLSM} result in the following modified 
harmonic function,
\begin{align}
H 
\ &= \ 
\frac{1}{g^2} 
+ \frac{1}{2} \sum_{a=1}^k \sum_{n_a = 1}^{\infty} \frac{\mathcal{N}}{R_a}
\left(
\e^{-n_a R_a \mathscr{R}^{\prime -1} + \i n_a \mathscr{R}^{\prime -1} \vartheta} 
+ \e^{-n_a R_a \mathscr{R}^{\prime -1} - \i n_a \mathscr{R}^{\prime -1} \vartheta}
\right) 
\nn \\
\ &= \ 
\frac{1}{g^2} + 
\sum_{a=1}^k \frac{\mathcal{N}}{2 R_a} 
\frac{\sinh (R_a \mathscr{R}^{\prime -1})}{\cosh (R_a
 \mathscr{R}^{\prime -1}) - \cos (\mathscr{R}^{\prime -1} \vartheta)}
\, .
\end{align}
We then compactify the $r^2$-direction with the small radius 
$\mathscr{R}$.
This procedure is equivalent to take 
$s_{1,a} = 2\pi \mathscr{R} a$ with $k \to \infty$ in the NLSM as we
have discussed before.
The summation over $a$ now becomes the continuous integral.
Again, the integral diverges and we introduce the cutoff 
$\Lambda$ and the renormalization scale $\mu$. 
Then the harmonic function of the $5^2_2$-brane of co-dimension two 
becomes 
\begin{align}
H \ = \ 
h'_0 + \frac{\mathcal{N}}{2\pi \mathscr{R}} 
\log \frac{\mu}{\varrho'} + \frac{\mathcal{N}}{2 \pi
\mathscr{R}} \sum_{n \neq 0} \e^{\i n \mathscr{R}^{\prime -1} \vartheta} K_0 (|n|
 \mathscr{R}^{\prime -1} \varrho')
\, ,
\label{eq:GLSM_harmonic}
\end{align}
where $\varrho' = \sqrt{(r^1)^2 + (r^3)^2}$ and $h'_0 = \frac{1}{g^2} +
\frac{\mathcal{N}}{2 \pi \mathscr{R}} \log \frac{4\pi \Lambda}{\mu}$.
Now we identify the fields in the NLSM and the coordinates of the geometry
$(r^1, y^2 ,r^3) = (X^1, \wt{X}^8, X^2)$ and $\varrho' = \varrho$.
Consequently we have the correspondences $\mathscr{R} = \mathcal{R}_8$,
%$\mathcal{N }= \mathscr{R}' = \mathcal{R}_9$, $h_0' = h_0$.
$\mathscr{R}' = \mathcal{R}_9$, $\mathcal{N} = \wt{R}_9$.
Then the result 
\eqref{eq:GLSM_harmonic} completely coincides with the one with the $X^9$-instanton
corrections elucidated in \eqref{eq:522harmonic}.

%%%%%%%%%%%%%%%%%%%%%%%%%%%%%%%%%%%%%%%%%%%%%%%%%%%%%%%%%%%%%%%%%%%%%%
\section{Conclusion and discussions}
\label{sect-conclusion}
In this paper we investigated the worldsheet instanton corrections to
the exotic $5_2^2$-brane geometry.
First we briefly reviewed the H-monopoles, 
the KK-monopoles and the $5_2^2$-brane from the spacetime viewpoint. 
These objects are related by the T-duality chain in type II string
theories. When the T-duality circle has finite radius, the geometry of
the H-monopoles is modified. This modification is interpreted as the
worldsheet instanton corrections which are analyzed in the language of
the GLSM \cite{Tong:2002rq}. 
We then mentioned 
the instanton corrections to the KK-monopoles
which causes the dependence on the winding coordinate in the harmonic
function \cite{Harvey:2005ab}. This results in the notion of the localized
KK-monopoles. 

In the main part of the current paper, we investigated the instanton
corrections to the geometry of the exotic $5_2^2$-brane from the two
distinct perspectives.
One is from the T-duality transformation of the localized KK-monopoles.
Since the localized KK-monopoles possess
the $U(1)$ isometry along the 
$X^8$-direction, we applied the Buscher rule to the geometry and 
obtained the $X^9$-instanton corrected $5^2_2$-brane geometry. 
We found that the correction to the harmonic function depends on the
winding coordinate $\vartheta$. 
The other is from the direct calculation
of the instantons in the GLSM for the multi-centered $k$
$\wh{5}^2_2$-branes  \cite{Kimura:2013fda}.
We found that the truncated model of the GLSM in the limit $g \to 0$ is
just the multi-centered generalization of the one discussed in
\cite{Tong:2002rq, Harvey:2005ab}.
The BPS solutions with the topological numbers 
$n_a = - \frac{1}{2\pi} \int \! \d^2 x \ F_{12,a}$ 
are the ANO vortices. 
We found that the instanton corrections to the co-dimension three
$\wh{5}^2_2$-branes are carried over to that of the $5^2_2$-brane geometry
by the appropriate choice of the FI parameters and the limit $k \to \infty$. 
The resulting harmonic function completely coincides with the one
obtained from the localized KK-monopoles.
The relation between the two results are summarized in Figure \ref{fig:instcorr}.
%%%%%%%%%%%%%%%%%%%%%%%%%%%
\begin{figure}[h]
\begin{center}
\includegraphics[scale=0.8]{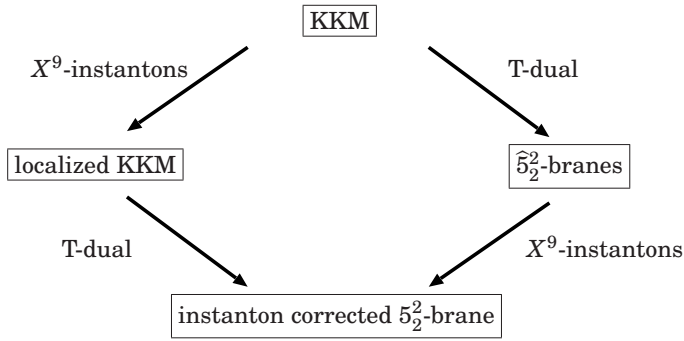}
\caption{Instanton corrections to the $5^2_2$-brane. 
Here KKM stands for the Kaluza-Klein monopole.}
\label{fig:instcorr}
\end{center}
\end{figure}
%%%%%%%%%%%%%%%%%%%%%%%%%%%

We note that the instanton corrections to the 
off-diagonal part of the metric and the vector $\omega_i$ are still
intractable. This is because the full geometry is no longer a solution
to conventional supergravities in the $g \to 0$ limit.
This can be understood from the harmonic function \eqref{eq:522h}. 
Even though the asymptotic radius of the T-dual circle is ill-defined
due to the renormalization scale $\mu$, we can go to the asymptotic
region $\varrho \sim \Lambda$ when the renormalization scale is large $\mu \sim
\Lambda$. 
We then find that the approximate radius of the T-dual circle is given by $g$.
Consequently, as we have mentioned in Section \ref{Tdual-chains}, the winding states become
massless in the limit $g\to 0$ and the supergravity approximation lose
its meaning.
The full solution of the $5^2_2$-brane 
with the $X^9$-instanton corrections would 
be studied in the framework of the double field theory \cite{Hull:2009mi} where 
the KK and the winding massless modes are treated democratically.

There is a question on the quantum deformation to the $5^2_2$-brane
geometry. We have studied the $X^9$-instanton effects which cause the
localization of the $5^2_2$-brane in the $\vartheta$-direction. 
This $\vartheta$-dependence originates from the T-dual of the
$X^9$-dependence in the KK-monopoles. The localization appears from the worldsheet
instanton effects or from the finite radius effect of the $X^9$ circle.
It is natural to consider the other quantum deformation caused by the
finiteness of the $X^8$ radius. Indeed, when we perform the discrete
summation over $l$ in \eqref{eq:X8X9sum} by keeping the radius
$\mathcal{R}_8$ finite, we have the harmonic function which depends not
only on $X^9$ but also on $X^8$. 
Even though the discrete summation in
\eqref{eq:X8X9sum} breaks the $U(1)$ isometry along 
the $X^8$-direction, 
we qualitatively expect that 
the harmonic function on the T-dualized $5^2_2$-brane geometry depends on the {\it two}
winding coordinates $\vartheta$ and 
$\vartheta' = X^8$.
The $\vartheta$-dependence is what we have investigated in this paper. 
The worldsheet origin of the $\vartheta'$-dependence is still an open
question. The localization in the 
$\vartheta'$-direction
is expected to be
induced by the topological term $\vartheta' F_{01}$ 
which is missing in our model. It is interesting to find the GLSM which contains
a topological term that enables one to localize the $5^2_2$-brane in the
$\vartheta'$-direction. We will study these issues in the next work.

%%%%%%%%%%%%%%%%%%%%%%%%%%%%%%%%%%%%%%%%%%%%%%%%%%%%%%%%%%
\section*{Acknowledgements}
The work of S.~S is supported in part by Sasakawa Scientific Research Grant from
The Japan Science Society and Kitasato University Research Grant for
Young Researchers.

}
\end{document}